# Observation of gigantic spin conversion anisotropy in bismuth

Naoki Fukumoto [1,$], Ryo Ohshima [1,$], Motomi Aoki [1], Yuki Fuseya [2],

Masayuki Matsushima [1], Ei Shigematsu [1], Teruya Shinjo [1], Yuichiro Ando [1,3],

Shoya Sakamoto [4], Masanobu Shiga [4], Shinji Miwa [4], and Masashi Shiraishi [1]

1. Department of Electronic Science and Engineering, Kyoto University, Kyoto, Kyoto 615-8510, Japan.

2. Department of Engineering Science, University of Electro-Communications, Chofu, Tokyo 182-8585, Japan.

3. PRESTO, Japan Science and Technology Agency, Honcho, Kawaguchi, Saitama 332-0012, Japan.

4. The Institute for Solid State Physics, The University of Tokyo, Kashiwa, Chiba 277-8581, Japan.

$ N.F. and R.O equally contributed to this work.

# Corresponding author: Ryo Ohshima(ohshima.ryo.2x@kyoto-u.ac.jp), Masashi Shiraishi (shiraishi.masashi.4w@kyoto-u.ac.jp)

**Abstract**

Whilst the *g*-factor can be anisotropic due to the spin-orbit interaction (SOI), its existence in solids cannot be simply asserted from a band structure, which hinders progress on studies from such the viewpoints. The *g*-factor in bismuth (Bi) is largely anisotropic; especially for holes at *T*-point, the *g*-factor perpendicular to the trigonal axis is negligibly small (< 0.112), whereas the *g*-factor along the trigonal axis is very large (62.7). We clarified in this work that the large *g*-factor anisotropy gives rise to the gigantic spin conversion anisotropy in Bi from experimental and theoretical approaches. Spin-torque ferromagnetic resonance was applied to estimate the spin conversion efficiency in rhombohedral (110) Bi to be 17%, which is unlike the negligibly small efficiency in Bi(111). Harmonic Hall measurements supports the large spin conversion efficiency in Bi(110). This is the first observation of gigantic spin conversion anisotropy as the clear manifestation of the *g*-factor anisotropy. Beyond the emblematic case of Bi, our study unveiled the significance of the *g*-factor anisotropy in condensed-matter physics and can pave a pathway toward establishing novel spin physics under *g*-factor control.

**INTRODUCTION**

Physical properties of solids can be largely dependent on crystal axis. A good example is the electric conductivity, the second rank tensor, which can be anisotropic in a crystal with low crystal inversion symmetry. Such the anisotropy is determined by the anisotropy in energy eigenvalues, i.e., an energy band structure. The spin Hall conductivity, which plays a dominant role in modern spin physics such as spintronics, spin-orbitronics, and spin conversion physics, is the third rank tensor. Since the additional rank comes from the spin direction, the spin Hall conductivity couples with the anisotropy of spin magnetic moment, i.e., the *g*-factor. Since the *g*-factor anisotropy is attributed to the spin-orbit interaction (SOI) and sensitive to description of the wave function [1], the existence of the *g*-factor anisotropy cannot be simply asserted from the anisotropy of the Fermi surface in the band structure and enables exploring novel unprecedented spin-related physical phenomena. Whilst the *g*-factor anisotropy is quite small in weak SOI elements (1.06 in graphite [2], 1.0006 in silicon [3], and 2.2 in strained Ge [4]), the anisotropy is gigantic in bismuth (Bi) that is a safe element with the largest SOI (14.4 for electron at the *L*-point and 560 for hole at the *T*-point [5–9]). Considering the coupling between the spin Hall conductivity and *g*-factor, spin conversion properties in Bi can be gigantically dependent on the crystal orientation.

The SOI plays a pivotal role in condensed matter physics and has attracted broad attention in view of spin-momentum locking in topological quantum materials [10–12], spin-charge interconversion in solids [13,14], spin manipulation in inorganic semiconductors [15–18], *etc*. Among the SOI-originated spin functions, spin-charge interconversion using the SOI is now attracting many physicists to accelerate the understandings of fundamental spin physics in solids. A representative spin-charge interconversion effect attributed to the SOI is the inverse spin Hall effect (ISHE) [19,20], the reciprocal effect of the spin Hall effect (SHE) [16,17]. The spin Hall angle $\theta_{SH}$ is an index of the spin conversion efficiency and is regarded to depend on the magnitude of the SOI. The strength of the SOI is roughly proportional to the fourth power of atomic number $Z$ [21,22], and the magnitudes of the $\theta_{SH}$ of heavy elements such as $\beta$-Ta [23], $\beta$-W [24], and Pt [25] are estimated to be 0.12, 0.33 and 0.08, respectively. Here, the magnitude $\zeta$ of the SOI for the 5$d$ electrons in Pt was calculated to be 46.1 mRy [26,27]. If the SOI of a material is the sole factor determining the spin conversion efficiency, Bi ($\zeta \sim 106.8$ mRy for the 6$p$ electrons) could exhibit the highest efficiency, resulting in the largest $\theta_{SH}$. However, spin conversion in Bi was quite small ($\theta_{SH} \ll 0.1$) in previous studies using rhombohedral (111) and polycrystalline Bi [28–33]. Given that Bi inherently possesses substantial SOI strength, the reason for the negligibly small $\theta_{SH}$ in the previous studies is quite unclear, which is one of the biggest mysteries in solid-state physics.

Here, in this work, we shed light on the spin conversion physics of rhombohedral (110) Bi (Bi(110)$_R$), which has been a less explored structure of Bi. In the case where the relativistic correction is relevant, the spin current is expressed as the product of the velocity times the spin magnetic momentum (e.g. Eq. (90) in ref. [34] and the other studies [35–37]). Therefore, the anisotropy of spin magnetic moment, or $g$-factor, yields the anisotropy of the spin Hall conductivity. Contrary to rhombohedral (111) Bi (Bi(111)$_R$), a gigantic spin conversion efficiency can be expected in Bi(110)$_R$ because its semimetallicity and large $g$-factor help generation of high spin polarization in plane of Bi film, which cannot occur in Bi(111)$_R$ where possible electron evacuation in thin films takes place [38] and the $g$-factor of holes can have a finite component only in the trigonal (Bi(111)$_R$) direction [5–7] (see Figs. 1(a) and 1(b)). An accomplished method to estimate the spin Hall angle of solids, spin-torque ferromagnetic resonance (ST-FMR), was applied, and the effect of the self-induced spin-orbit torque (SOT) [39–43] in an adjacent ferromagnet of the Bi(110)$_R$ was taken into account to obtain compelling results enabling precise estimation of the $\theta_{SH}$ of Bi(110)$_R$. The $\theta_{SH}$ of Bi(110)$_R$ is estimated to be 0.17, although that of Bi(111)$_R$ was quite small [28–33]. Harmonic Hall measurements provided steadfast supporting evidence of this gigantic efficiency.

**SAMPLE STRUCTURE AND MEASUREMENT SETUP**

An epitaxial Ni(5 nm)/Bi($t_{Bi}$) film was grown on a single-crystal MgO(001) substrate by molecular beam epitaxy. The Bi layer, the thickness $t_{Bi}$ of which ranged from 0 to 12 nm with a wedged structure, was grown with substrate cooling (< 250 K) to obtain a (110)-oriented rhombohedral epitaxial film [44] (see Fig. 1(c) and Appendix A for details). Figure 1(d) shows a $\theta$-$2\theta$ X-ray diffraction profile obtained from the Ni/Bi film. A clear face-centered cubic (fcc) 002 Ni peak and a rhombohedral 110 Bi peak were observed, indicating that highly oriented Bi(110)$_R$ was grown on single-crystal Ni, as expected from the *in situ* reflection high energy diffraction (RHEED) measurement during the growth process (see Appendix B for the RHEED spectra). We also prepared Ni/Bi and Fe/Bi samples with and without substrate cooling during the Bi growth for control experiments. Unless otherwise noted, the Ni/Bi film with substrate cooling is used in the main text. The Ni/Bi film was formed into 10 μm × 65 μm rectangular channels and connected to Ti(3 nm)/Au(150 nm) electrodes by using Ar-ion milling, electron-beam (EB) lithography, and EB evaporation (See Sec. 1 of Supplemental Material [45] for the effect of baking on the Ni/Bi interface). For the ST-FMR measurement, the electrode was fabricated as the channel in the form of the signal lines of shorted coplanar waveguides. Here, the longitudinal direction of the Ni/Bi channel was designed to be in the Ni[010] direction. The ST-FMR measurement [25,46] was carried out by injecting an rf current with an input power of 10 mW and applying an external magnetic field along $\phi = 45°$ and $\phi = 225°$ with respect to the longitudinal direction of the Ni/Bi channel (see Fig. 2(a)). An Oersted field and a spin current from Bi produced by the SHE induce magnetization precession in Ni. Then, the resistance of the channel oscillates due to the anisotropic magnetoresistance, resulting in the DC voltage $V_{\text{DC}}$ as a spin-torque diode effect [46] (see Appendix A for details).

**ST-FMR MEASUREMENT AND LARGE SPIN-CHARGE CONVERSION IN BISMUTH**

Figure 2(b) shows the Bi thickness dependence of the ST-FMR spectra. The frequency was set to 7 GHz in this experiment. The ST-FMR signal can be deconvoluted into symmetric and antisymmetric Lorentzian signals by fitting the following equation [25]:

$$V_{\text{DC}} = -C\left(S\frac{\Delta^2}{\mu_0^2(H-H_{\text{res}})^2+\Delta^2} + A\frac{\Delta\mu_0(H-H_{\text{res}})}{\mu_0^2(H-H_{\text{res}})^2+\Delta^2}\right) + a\mu_0 H + b, \qquad (1)$$

where $C$ is a coefficient concerning the anisotropic magnetoresistance, $S$ is the symmetric component, $A$ is the antisymmetric component, $\mu_0$ is the vacuum permeability, $H_{\text{res}}$ is the resonance field, $\Delta$ is the half-width at half-maximum of the spectrum, and $a$ and $b$ are coefficients of the linear background. Here, $H_{\text{res}}$ and $\Delta$ are related to the characteristics of the magnetization dynamics and are evaluated through the frequency dependence. Since we used single-crystal fcc (001) Ni as the ferromagnetic layer, the Kittel equation concerning the magnetocrystalline anisotropy was used to fit the frequency dependence of $H_{\text{res}}$ (see Sec. 3 of Supplemental Material [45] for the full expression of the Kittel equation) [47]. The frequency $f$ dependence of $\Delta$ can be fitted with the following equation [48]: $\Delta = \Delta_0 + 2\pi\alpha f/\gamma$, where $\Delta_0$ is the frequency-independent scattering parameter, $\gamma$ is the gyromagnetic ratio, and $\alpha$ is the Gilbert damping parameter. Both frequency dependences were well fitted, and thus, we can discuss the magnetic properties based on the ST-FMR measurement results (see Sec. 4 of Supplemental Material [45] for the frequency dependence of the ST-FMR signals and their evaluations). The Gilbert damping parameter increases with increasing Bi thickness and starts to be saturated at approximately $t_{Bi}$ = 6 nm (see Fig. 2(c)). Since the Gilbert damping parameter depends on the spin torque induced by the injection of the spin current from Bi via the SHE, it depends on the spin diffusion length of Bi [49]. Thus, this result indicates that we observed the SHE in bulk Bi, not interfacial effects, and more importantly, the saturation of the damping parameter at approximately 6 nm is a manifestation of the short spin diffusion length in Bi(110)$_R$.

The spin-charge conversion efficiency, $\eta$, is in principle estimated from the ratio of $S$ and $A$ (see Sec. 4 of Supplemental Material [45] for details). The Bi thickness dependence of $\eta$ is shown in Fig. 2(c), where the dependence is quite similar to that of $\alpha$. For precise estimation of the efficiency in ST-FMR, the self-induced SOT in Ni is taken into account as follows. Recently, the SHE occurring in a ferromagnet itself has been receiving attention [50–55]. Since the SHE in a ferromagnet provides a substantial contribution in ST-FMR, the SOT induced by the spin current generated in the ferromagnet can be superimposed in the ST-FMR measurement [39–43]. In particular, when the nonmagnetic layer of the ST-FMR device is highly resistive, most of the rf current is injected into the ferromagnetic layer, and the SOT generated in the ferromagnetic layer becomes large, resulting in overestimation of the spin-charge conversion efficiency of the nonmagnetic layer [43]. Here, the typical conductivities of Bi and Ni are $2.4 \times 10^5$ S/m [56] and $8.0 \times 10^6$ S/m [54,57], respectively, so the SOT from Ni should be considered for precise estimation of the spin-charge conversion in Bi. The red solid line in Fig. 2(c) shows the fitting result of the thickness dependence of $\eta$ considering the self-induced SOT in Ni (see Sec. 5 Supplemental Material [45] for the detailed parameters in the calculation). $\theta_{Bi}$ = 0.17 allows the best fit in reproducing our experimental results, and the Bi layer is responsible for the spin-charge conversion in the Ni/Bi channels despite its high resistivity (notably, $\theta_{Bi}$ can be augmented when the conductivity of Bi increases, i.e., the estimated efficiency here is a kind of lower limit; see also Sec. 5 of Supplemental Material [45]).

The efficiency in Bi(110)$_R$ is comparable to that in heavy metals such as Pt and $\beta$-W, which means realization of one of the largest spin conversion efficiencies in solids. More importantly, a substantially large spin-charge conversion efficiency can be achieved in Bi(110)$_R$, in sharp contrast to Bi(111)$_R$ [28,29].

**CRYSTAL-ORIENTATION ANISOTROPY OF SPIN-CHARGE CONVERSION IN BISMUTH**

For the Fe/Bi samples, the substrate cooling changed crystal orientation of the Bi, Bi(110)$_R$ from the sample with cooling and Bi(111)$_R$ from the sample without cooling, as indicated by the XRD patterns (see Figs. 3(a) and 3(b)). Although XRD patterns were measured at different conditions and a peak from MgO looks different from each other, clear difference of Bi peaks due to their crystallinity was observed. We carried out the ST-FMR measurements on the Fe(6 nm)/Bi(10 nm) samples with different crystal structures as shown in Figs. 3(c) and 3(d). The ST-FMR signal strongly depends on the crystal orientation of Bi; the signal obtained from Fe/Bi(111)$_R$ exhibits almost only anti-symmetric component and the symmetric component attributed to the spin-orbit torque was almost missing. We estimated the spin-charge conversion (SCC) efficiency $\eta$ to be 0.34 for Fe/Bi(110)$_R$ and $-0.027$ for Fe/Bi(111)$_R$, respectively. As the previous study, Bi(111)$_R$ showed the quite small SCC efficiency. On the other hand, we observed the gigantic SCC in Bi(110)$_R$ as demonstrated in Ni/Bi(110)$_R$ as described above, indicating that crystal orientation strongly affects the SCC in Bi. By the combination of these experiments and previous reports claimed that small SCC on Bi, especially for Bi(111)$_R$, we can say that the SCC on Bi possesses a large anisotropic SCC efficiency for Bi(110)$_R$ and Bi(111)$_R$.

**HARMONIC HALL MEASUREMENT**

To confirm the large spin-charge conversion efficiency in Bi(110)$_R$, we carried out a harmonic Hall measurement [58–60]. The Hall bar structure was fabricated from a Ni/Bi film, where the Bi thickness was 11 nm. The harmonic Hall measurement was carried out by injecting an rf current with an effective amplitude of 5 mA and applying an external magnetic field rotating in the plane of the Ni/Bi interface ($\phi = 0° – 360°$). The second harmonic Hall voltage $V_{2\omega}$ was measured with a lock-in amplifier (see Fig. 4(a) and Sec. 6 of Supplemental Material [45] for details). Figure 4(b) shows the $\phi$ dependence of $V_{2\omega}$ with $\mu_0 H = 0.2$ T and $\mu_0 H = 4.0$ T (see also Sec. 6 of Supplemental Material [45] for the result with other magnetic fields). $V_{2\omega}$ was fitted with the following equation [60–62]:

$$
\begin{aligned}
V_{2\omega} &= A(H) \cos \phi + B(H) \cos 2\phi \cos \phi \\
&= -\frac{1}{2}\left(V_{\text{AHE}} \frac{H_{\text{DL}}}{H_{\text{K}} + H} + V_{\text{ANE}} + V_{\text{ONE}} H\right) \cos \phi \\
&\quad - V_{\text{PHE}} \frac{H_{\text{FL}} + H_{\text{Oe}}}{H} \cos 2\phi \cos \phi,
\end{aligned}
\tag{2}
$$

where $H_{\text{DL}}$ and $H_{\text{FL}}$ are the damping-like torque effective field and the field-like torque effective field, respectively, $H_{\text{Oe}}$ is the Oersted field, $H_{\text{K}}$ is the out-of-plane anisotropy field, $V_{\text{ANE}}$ is the anomalous Nernst voltage, and $V_{\text{ONE}}$ is the ordinary Nernst voltage. $V_{\text{AHE}} = 2.50$ mV and $V_{\text{PHE}} = -0.72$ mV are the anomalous Hall voltage and planar Hall voltage obtained from first harmonic voltage $V_{1\omega}$ (see Sec. 6 of Supplemental Material [45] for details of the harmonic Hall measurement and fitting parameters). Figure 4(c) shows the magnetic field $H$ dependence of the fitting parameter, $A(H)$. $A(H)$ can be deconvoluted into three components ($V_{\text{AHE}}$, $V_{\text{ANE}}$, and $V_{\text{ONE}}$ as shown in Eq. (2)), and we can derive $\mu_0 H_{\text{DL}} = 160$ μT from the $V_{\text{AHE}}$ component. From the magnetic field dependence of $B(H)$, $\mu_0(H_{\text{FL}} + H_{\text{Oe}})$ was estimated to be 29 μT (Fig. 4(d)). Here, $\mu_0 H_{\text{Oe}} = \mu_0 I_{\text{Bi}}/2w$ was estimated to be 28 μT, where $I_{\text{Bi}}$ is the electric current in Bi. The damping-like torque efficiency $\xi_{DL}$ and field-like torque efficiency $\xi_{FL}$ can be described with $H_{DL}$ and $H_{FL}$ as follows [63]:

$$\xi_{DL(FL)} = \left(\frac{2e}{\hbar}\right)\mu_0 M_S t_{Ni} t_{Bi} w \frac{H_{DL(FL)}}{I_{Bi}}, \quad (3)$$

where $w$ is the width of the Ni/Bi channel. $\xi_{DL}$ and $\xi_{FL}$ were estimated to be $2.7 \times 10^{-1}$ and $2.4 \times 10^{-3}$, respectively. We postulate transparency at the Ni/Bi interface, and $\theta_{Bi} \sim \xi_{DL} = 0.27$ was obtained from the harmonic Hall measurement. The ratio $\xi_{FL}/\xi_{DL}$ was small, estimated to be $8.7 \times 10^{-3}$ from the harmonic Hall measurement, which is very similar to the results of the ST-FMR measurement (see Sec. 6 of Supplemental Material [45]). Note that the difference in $\theta_{Bi}$ obtained from the different measurements originated from the difference in the effects considered in the analysis except for the SHE in Bi: the ST-FMR analysis considered the self-induced SOT in the Ni layer, and the harmonic Hall measurement analysis included thermal effects such as the anomalous Nernst effect (ANE) in Ni and the ordinary Nernst effect (ONE) in Bi. Despite such differences, both measurements indicated a large spin-charge conversion efficiency in Bi. Thus, the result obtained from the harmonic Hall measurement strongly supports our claim that a gigantic spin-charge conversion efficiency was achieved in our Bi(110)$_R$, obtained from ST-FMR.

**DISCUSSION**

Considering the gigantic $\theta_{Bi}$ (= 0.17) in our Bi(110)$_R$ and the negligibly small $\theta_{Bi}$ in Bi(111)$_R$ (see Fig. 3 for control experiments on Fe/Bi(111)$_R$ and Fe/Bi(110)$_R$. See also our previous study [28] and the study by the other group [29].), the anisotropy of physical properties in Bi should shed light on. Here, we emphasize that while the gigantic $\theta_{Bi}$ can inherently originate from the large atomic SOI of Bi, the large atomic SOI does not always rationalize the highly anisotropic $\theta_{Bi}$. Hence, we shed light on the anisotropic $g$-factor of Bi, which can couple with the spin Hall conductivity and determine spin conversion efficiency from a theoretical viewpoint to reveal the underlying physics of our findings.

The anisotropy of the magnetic moment (or the $g$-factor) in Bi has been well established both experimentally and theoretically [5–9]. In particular, the anisotropy of the $g$-factor of holes at the $T$-point is very large. The symmetry of the hole wave function ($T_{45}^-$) prohibits the $g$-factor from being oriented perpendicular to the trigonal direction [8] (see Figs. 1(a) and 1(b)). The $g$-factor of holes can have a finite component only in the trigonal direction, and indeed, this Ising-like $g$-factor has been experimentally confirmed [5–7]. Figure 5 shows the crystal structure and band structure of bulk Bi. The Bi $[111]_R$ direction corresponds to the trigonal axis. The thin Bi$(111)_R$ film consists of buckled honeycomb bilayers (BL). In contrast, the thin Bi$(110)_R$ film consists of almost flat quasi-square BL. In bulk Bi, electrons appear at the $L$-point, and holes appear at the $T$-point. The effective mass tensor $m_i$, cyclotron mass $m_c$, and $g$-factor $g$ for both electrons and holes determined experimentally are listed in Table 1 [7,9]. The spin splitting factor, given by the multiplication of the $g$-factor and the cyclotron mass as $M = gm_c/2$, expresses the SOI's impact. (Note that $M$ can be determined by the phase of the quantum oscillation.) Table 1 clearly shows that the $M$ for holes is highly anisotropic compared to that for electrons. The $g$-factor perpendicular to the trigonal axis is shown to be vanishingly small ($g_{\text{bin}} = g_{\text{bis}} < 0.112$ within the experimental accuracy), whereas the $g$-factor along the trigonal axis is very large ($g_{\text{tri}} = 62.7$). (see Appendix C for anisotropic $g$-factor due to the crystalline SOI).

The spin current is strongly restricted by this directivity of the hole $g$-factor. In the SHE, the spin current flows in the direction perpendicular to both the electric current and the $g$-factor. For holes at the $T$-point of Bi, the spin current can flow only in the trigonal plane, i.e. in the plane parallel to the $(111)_R$ interface. Consequently, the spin current is prohibited from being injected through the Bi$(111)_R$ interface. This can be the reason for the vanishingly small $\theta_{\text{Bi}}$ in Bi(111)$_R$. Meanwhile, the spin current can flow through the Bi(110)$_R$ interface, taking advantage of the large *g*-factor. Note that contributions from the electrons at the *L*-point have not been detected in Bi(111)$_R$ structures, which is ascribable to the evacuation of bulk electrons caused by the remarkable size effect with the tiny effective mass ($m_{tri}$ = 0.00585). In fact, such evacuation of bulk electrons has been observed for ultrathin Bi(111)$_R$ films (below 30 bilayers (30BL), less than 11.7 nm) [38]. The previous (up to 10BL [64]) and our present first-principles (up to 15BL) calculations also support bulk electron evacuation in ultrathin Bi(111)$_R$ films. Figure 6 shows the two-dimensional (2D) BZ for Bi(111)$_R$ film and its band structures for 7BL and 15BL. Density functional theory (DFT) calculations were performed using BAND software of Amsterdam Modeling Suite [65,66] (see Appendix A for details). For the Bi(111)$_R$ surface, the $\bar{\Gamma}$-point in the 2D BZ corresponds to the Γ- and *T*-points in the bulk 3D BZ, where holes appear. The $\bar{M}$-point in 2D corresponds to the *L*-point in 3D, where electrons appear. The bulk valence bands cross the Fermi energy at approximately the $\bar{\Gamma}$-point in the films with ≥ 7BL. In contrast, the bulk conduction bands ($\bar{M}$-point) do not cross the Fermi energy in the films with ≤ 13BL. These results agree well with the previous calculations below 10BL [64]. Even in the 15BL film (~ 6 nm), the bulk electron density is extremely low, although the bulk conduction band crosses the Fermi energy. This evacuation of bulk electrons is consistent with the experimental observation by Hirahara et al. [38]. (see Sec. 9 of Supplemental Material [45] for the results of the first principles calculations for various Bi thickness and Bi(110)$_R$ film). Thus, the largely anisotropic *g*-factor of holes for Bi(111)$_R$ and Bi(110)$_R$, and possible evacuation of electrons in Bi(111)$_R$ can rationalize the substantially large spin conversion efficiency in Bi(110)$_R$ and the absence of spin conversion in Bi(111)$_R$.

**CONCLUSION**

We observed gigantic spin conversion anisotropy in Bi. Unlike negligibly small efficiency in Bi(111)$_R$, Bi(110)$_R$ exhibits large spin-charge conversion of 17%. This long-term-awaited large efficiency stems from the coexistence of the anisotropic *g*-factor and semimetalllic nature of Bi(110)$_R$. The theoretical consideration focusing on the large *g*-factor anisotropy in Bi enables reconciliation of the large SOI in Bi and missing spin conversion in Bi(111)$_R$. Although it has been widely recognized that the *g*-factor can be anisotropic due to the crystal SOI since more than half century ago, spin conversion utilizing the *g*-factor anisotropy has been unexplored. This work unveiled the significance of the *g*-factor anisotropy in condensed-matter physics. Furthermore, beyond the emblematic case of Bi, our findings can pave an avenue toward establishing novel spin physics under *g*-factor control.

**Acknowledgements**

This research is supported in part by the Japan Society for the Promotion of Science (JSPS) Grant-in-Aid for Scientific Research (S) (No. 16H06330), JSPS Grant-in-Aid for Scientific Research (B) (No. 19H01850), JSPS Grant-in-aid for Scientific Research (A) (No. 22H00290), and the Spintronics Research Network of Japan (Spin-RNJ). A part of this work was supported by the Kyoto University Nano Technology Hub through the "Nanotechnology Platform Project" sponsored by the Ministry of Education, Culture, Sports, Science and Technology (MEXT), Japan.

**Appendix A: DETAILED EXPERIEMNTAL SETUP**

1. **Sample fabrication**

An epitaxial Ni(5 nm)/Bi($t_{Bi}$) film was grown on a single-crystal MgO(001) substrate by molecular beam epitaxy. First, the MgO substrate was annealed at 800°C for 10 min. The 5 nm Ni layer was deposited at room temperature and postannealed at 350°C for 15 min. The Bi layer, the thickness $t_{Bi}$ of which ranged from 0 to 12 nm with a wedged structure, was grown with substrate cooling (< 250 K) to obtain a (110)-oriented rhombohedral epitaxial film [44]. The multilayer was capped with a MgO(5 nm)/SiO$_2$(5 nm) layer to avoid surface oxidation. The Ni/Bi film was formed into 10 μm × 65 μm rectangular channels and connected to Ti(3 nm)/Au(150 nm) electrodes by using Ar-ion milling, EB lithography, and EB evaporation. For the ST-FMR measurement, the electrode was fabricated as the channel in the form of the signal lines of shorted coplanar waveguides. In contrast, for the harmonic Hall measurement, the channel was formed in the typical Hall bar structure, where the distance between two terminals was 35 μm and the width of the terminals were 5 μm. Here, the longitudinal direction of the Ni/Bi channel was designed to be in the Ni[010] direction. For EB lithography, we used a charge-dissipating agent (ESPACER 300Z, Showa Denko) on the EB lithography resist to avoid charge-up with the insulative substrate. To make ohmic contact on the channel with the electrode, Ar-ion milling in a load-lock chamber was performed to remove the capping layer prior to electrode deposition.

2. **Measurement setup**

The ST-FMR measurement [25,46] was carried out by injecting an rf current with an input power of 10 mW produced by a signal generator and applying an external magnetic field along 45° and 225° with respect to the longitudinal direction of the Ni/Bi channel (see Fig. 2(a)). The DC voltage $V_{DC}$ induced by the ferromagnetic resonance of the Ni layer was measured with a bias-T and a lock-in amplifier. The rf current frequency ranged from 7 to 12 GHz. For the Bi thickness dependence measurement, the frequency was set to 7 GHz. All measurements were carried out at room temperature.

The harmonic Hall measurement [58–60] was carried out by injecting an rf current with an effective amplitude of 5 mA and a frequency of 17 Hz produced by the internal source meter of a lock-in amplifier and applying an external magnetic field produced by a Physical Property Measurement System (Quantum Design) (see Fig. 4(a)). The magnetic field ranged from 0.2 mT to 4.0 T. The 2nd harmonic Hall voltage $V_{2\omega}$ was measured by the lock-in amplifier with the magnetic field rotating in the plane of the Ni/Bi interface ($\phi$ = 0° – 360°). For estimation of the fitting parameters, 1st harmonic measurements with an out-of-plane magnetic field (anomalous Hall measurement) and with an in-plane $\phi$ = 0° – 360° (planar Hall measurement) were carried out. All measurements were carried out at room temperature.

3, Density functional theory calculations

Density functional theory (DFT) calculations were performed using BAND software of Amsterdam Modeling Suite [65,66]. We used the fast inertial relaxation engine (FIRE) for geometry optimization [67]. In the generalized gradient approximation (GGA) of DFT, the Perdew-Burke-Ernzerhof (PBE) exchange-correlation functional was used [68] together with double-zeta polarized basis sets and numerical orbitals with a large frozen core. The relativistic effects were taken into account by the noncollinear method.

**APPENDIX B: IN SITU RHEED PATTERNS DURING THE FILM GROWTH**

Figure 7 shows the *in situ* RHEED patterns of the MgO substrate, Ni, and Bi surfaces during film growth, where the incident electron beam was parallel to the [100] azimuth of the MgO substrate. Sharp streak patterns were observed in the region of thick Bi films (> 6 nm). These observations demonstrate that a sharp heteroepitaxial interface was obtained at Ni/Bi.

While the spotty patterns observed in the thick Bi region show that the Bi surface was not atomically flat, these results show that all the layers were epitaxially grown. The RHEED pattern for Bi is consistent with the X-ray diffraction pattern in Fig. 1(d) and with a previous study [44] showing a rhombohedral (110) Bi (Bi(110)$_R$) structure.

**APPENDIX C: ANISOTROPIC G-FACTOR DUE TO THE CRYSTALLINE SOI**

The general formula of the *g*-factor is given in the following form:

$$g_i = 2m\sqrt{G_i}, \tag{C1}$$

$$G_i = 4\left|\left(\sum_{n\neq 0}\frac{\boldsymbol{t}_n \times \boldsymbol{u}_n}{E_0 - E_n}\right)_i\right|^2 - \left(\sum_{n\neq 0}\frac{\boldsymbol{t}_n \times \boldsymbol{t}_n^* + \boldsymbol{u}_n \times \boldsymbol{u}_n^*}{E_0 - E_n}\right)_i^2, \tag{C2}$$

where *m* is the electron mass and $E_n$ is the energy of the *n*-th band (*n* = 0 is the band under consideration). $t_n$ and $u_n$ are the interband matrix elements of the velocity operator between the 0-th and *n*-th bands for the same spins and opposite spins, respectively. The hole band (the top valence band) in Bi has $T_{45}^-$ symmetry (cf. Fig. 5). It has the matrix elements only with $T_{45}^+$ and $T_6^-$ as follows [8]:

$$\boldsymbol{t}_n^{(6)} = \langle T_{45}^-(1)|\boldsymbol{v}|T_6^+(n)\rangle = (-a_n, ia_n, 0), \tag{C3}$$

$$\boldsymbol{u}_n^{(6)} = \langle T_{45}^-(1)|\boldsymbol{v}|CT_6^+(n)\rangle = (-a_n, -ia_n, 0), \tag{C4}$$

$$\boldsymbol{t}_n^{(45)} = \langle T_{45}^-(1)|\boldsymbol{v}|T_{45}^+(n)\rangle = (0,0,b_n), \tag{C5}$$

where $a_n$ and $b_n$ are complex numbers and operator *C* is the product of space inversion and time-reversal operators, i.e., $|T_6^+(n)\rangle$ and $|CT_6^+(n)\rangle$ are the Kramers doublet. The *x*-, *y*-, and *z*-directions are taken to be the binary, bisectrix, and trigonal directions, respectively. From Eq.

(C2), the *g*-factor is given by the outer products of $t_n$ and $u_n$. Therefore, only $G_z$ has a finite component (cf. Eqs. (C3-C5)). In contrast, $G_{x,y}$ is zero because $(t_n \times u_n)_{x,y} = 0$. Consequently, the *g*-factor is finite only when the magnetic field is parallel to the trigonal direction (Bi[111]$_R$ direction), and it is zero when the field is perpendicular to the trigonal direction. A schematic view of the highly anisotropic *g*-factor for holes is given in Fig. 1(a) and 1(b).

analysis of the ST-FMR and harmonic Hall measurements.

**Figure captions**

Fig. 1 Schematic of the relationship between the anisotropic *g*-factor of Bi and generation of a spin current via SHE $J_s$ for (a) Bi(110)$_R$ (in this study) and (b) Bi(111)$_R$. Here, the subscript R indicates a rhombohedral structure. (c) Schematic of Bi(110)$_R$ structure and the Bi(110)$_R$ plane view from the Bi[110]$_R$ direction. (d) X-ray diffraction pattern (Cu-K$\alpha$ radiation) of the Ni/Bi film. Peaks at 38.5 and 43.1 deg. were obtained from the MgO(001) substrate in K$\beta$ radiation and K$\alpha$ radiation, respectively.

Fig. 2 (a) Schematic of the setup of the ST-FMR measurement. (b) ST-FMR signals obtained from the Ni/Bi($t_{Bi}$) channel at 7 GHz. Here, $t_{Bi}$ ranges from 2 to 12 nm. Solid lines indicate the fitting with Eq. (1) for each $t_{Bi}$. (c) Bi thickness dependence of the Gilbert damping parameter $\alpha$ (black squares) and the spin-charge conversion efficiency $\eta$ (red filled circles). $\alpha$ and $\eta$ increase with increasing Bi thickness and start to saturate at approximately 6 nm Bi thickness. The red solid line indicates the fitting result considering the self-induced SOT in Ni. The calculation well reproduces the data when $\theta_{Bi}$ is equal to 0.17.

Fig. 3 XRD spectra obtained from (a) Fe/Bi with (w/) substrate cooling, and (b) Fe/Bi without (w/o) substrate cooling during the Bi growth. ST-FMR signals obtained from (c) Fe/Bi(110)$_R$, with substrate cooling during the Bi growth, and (d) Fe/Bi(111)$_R$, without cooling.

Fig. 4 (a) Schematic of the setup of the harmonic Hall measurement. Here, the Bi thickness was 11 nm. (b) Azimuth angle $\phi$ dependence of the second harmonic Hall voltage $V_{2\omega}$ for various magnetic fields. Here, the magnetic field rotates in the plane of the Ni/Bi interface ($\phi = 0° - 360°$). Solid lines indicate the fitting with Eq. (2) in the main text. (c) Magnetic field dependence of the fitting parameter $A(H)$. $A(H)$ was deconvoluted into the anomalous Hall voltage (red line), ordinary Nernst effect in Bi (green line), and anomalous Nernst effect in Ni (blue line). The black line indicates the sum of the three components. (d) Magnetic field dependence of the fitting parameter $B(H)$. The solid line indicates the fitting with the definition of $B(H)$ in Eq. (2).

Fig. 5. (a) Crystal structure of Bi. The atoms of different colors belong to different bilayers. (b) Band structure calculated by Liu-Allen's tight-binding model [69]. $T_{45,6}$ indicates the symmetry of the wave function at the $T$-point [8].

Fig. 6. (a) Brillouin zone for Bi(111)$_R$ and its band structures for (b) 7BL and (c) 15 BL by DFT. The horizontal dashed lines indicate the Fermi energy.

Fig. 7. *In situ* RHEED images of the surface of the MgO substrate, Ni, and Bi.

Table 1. Effective mass tensor $m_i$, cyclotron mass $m_c$, $g$-factor $g$, and spin splitting factor $M = gm_c/2$ for electrons and holes [7,9].

**Figures and Table**

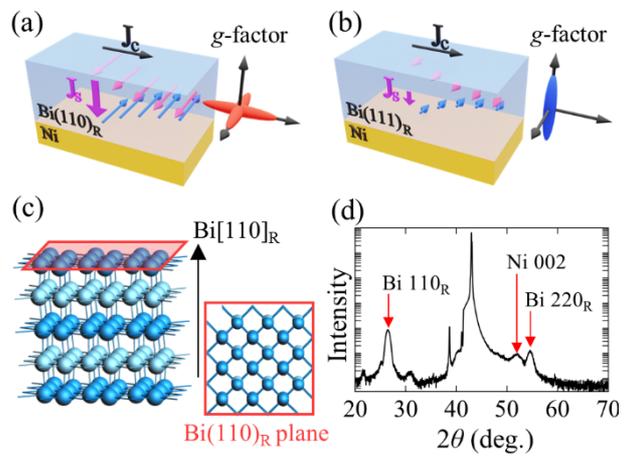

Fig. 1 N. Fukumoto et al.

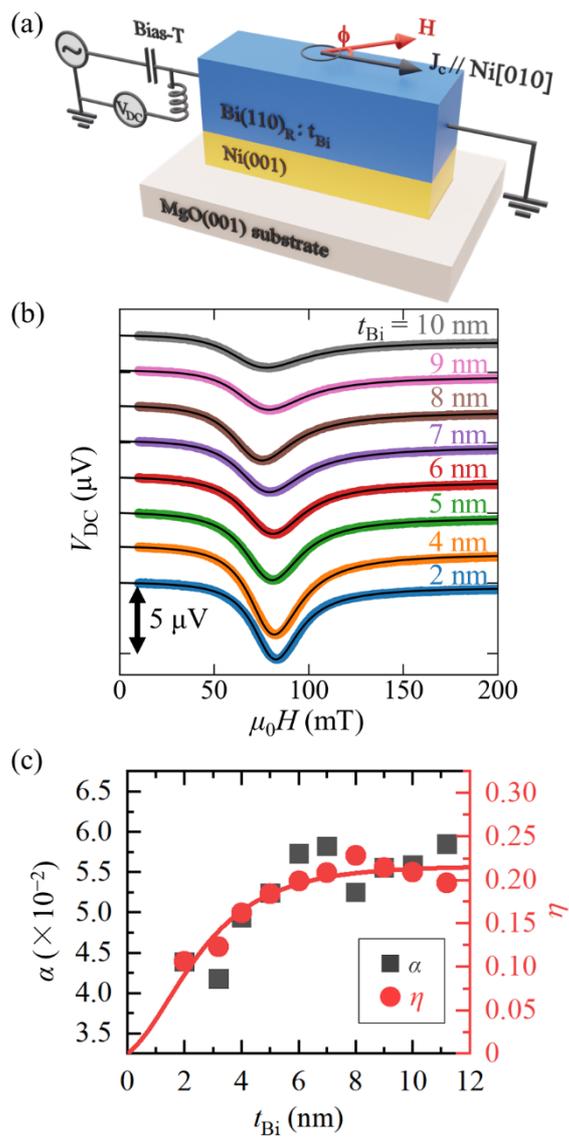

Fig. 2 N. Fukumoto et al.

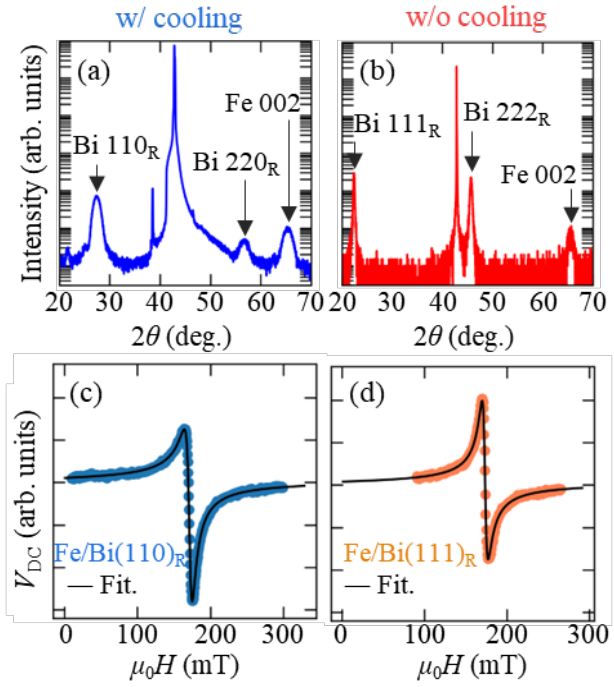

Fig. 3 N. Fukumoto et al.

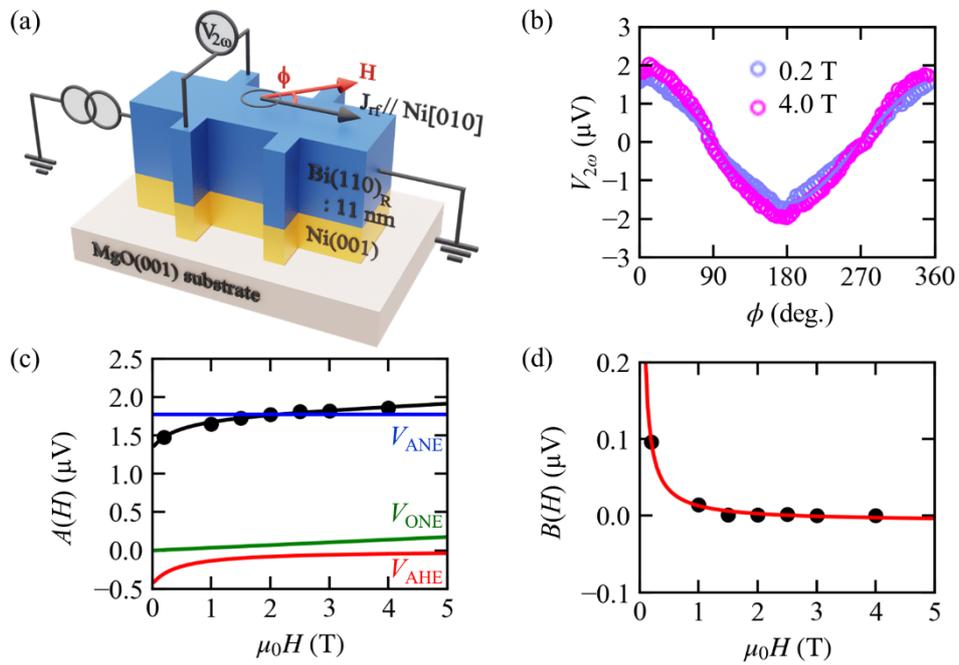

Fig. 4 N. Fukumoto et al.

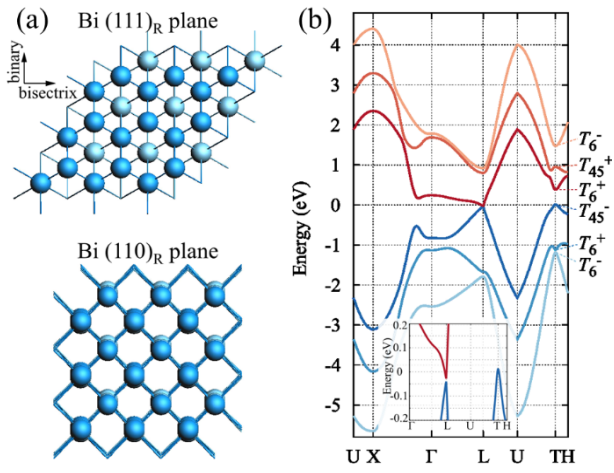

Fig. 5 N. Fukumoto et al.

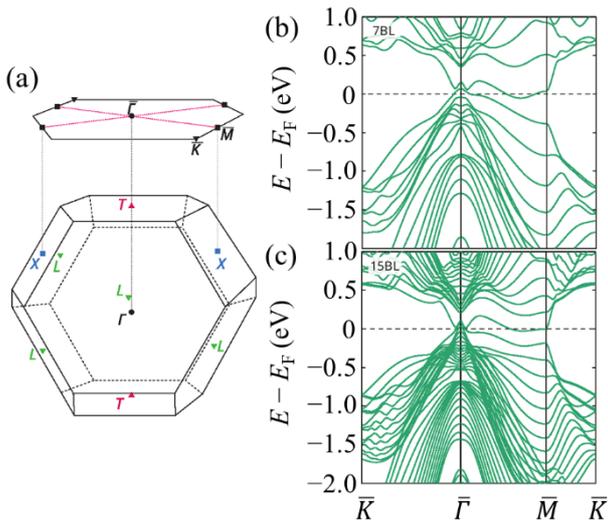

Fig. 6 N. Fukumoto et al.

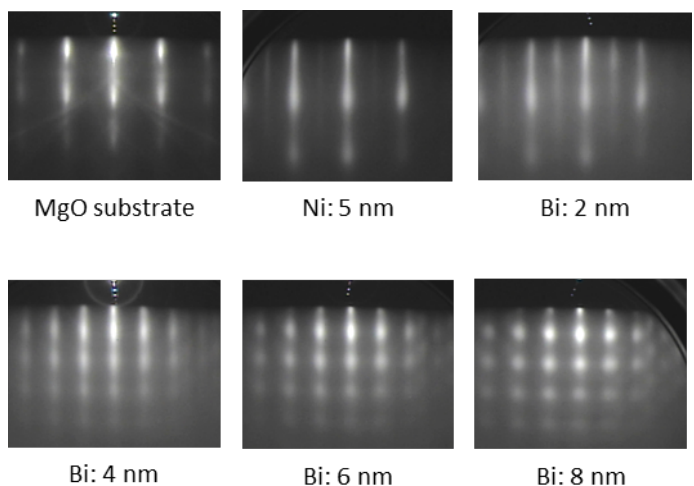

Fig. 7 N. Fukumoto et al.

Table 1 N. Fukumoto et al.

| | **For electrons** | | | |
|---|---|---|---|---|
| $i$ | 1 (binary) | 2 (bisectrix) | 3 (trigonal) | 4 |
| $m_i$ | 0.00124 | 0.257 | 0.00585 | −0.0277 |
| $m_c$ | 0.0272 | 0.00189 | 0.0125 | -- |
| $g$ | 66.2 | 1084 | 151 | -- |
| $M$ | 0.90 | 1.02 | 0.94 | -- |
| | **For holes** | | | |
| $i$ | 1 (binary) | 2 (bisectrix) | 3 (trigonal) | 4 |
| $m_i$ | 0.0678 | 0.0678 | 0.721 | 0 |
| $m_c$ | 0.221 | 0.221 | 0.0678 | -- |
| $g$ | 0.112 | 0.112 | 62.7 | -- |
| $M$ | 0.012 | 0.012 | 2.13 | -- |